\documentclass[%
 reprint,
superscriptaddress,
showpacs,preprintnumbers,
 amsmath,amssymb,
 aps,
]{revtex4-1}

\usepackage{xcolor}
\usepackage{graphicx}
\usepackage{dcolumn}
\usepackage{bm}

\usepackage[caption=false]{subfig}
\usepackage{epstopdf} 

\newcommand\redout{\bgroup\markoverwith{\textcolor{red}{\rule[.5ex]{2pt}{0.4pt}}}\ULon}

\usepackage[colorlinks=true,citecolor=blue]{hyperref}
\hypersetup{colorlinks=true,citecolor=blue,linkcolor=red,urlcolor=blue}

\begin{document}
\title{Quasiparticle band-gap renormalization in doped monolayer MoS$_2$}
\author{Azadeh Faridi}
\email{azadeh.faridi@ipm.ir}
\affiliation{School of Physics, Institute for Research in Fundamental Sciences (IPM), Tehran 19395-5531, Iran}
\author{Dimitrie Culcer}
\affiliation{School  of  Physics,  University  of  New  South  Wales,  Kensington,  NSW  2052,  Australia}
\affiliation{ARC Centre of Excellence in Future Low-Energy Electronics Technologies,  UNSW Node,  Sydney 2052,  Australia}
\author{Reza Asgari}
\email{asgari@ipm.ir}
\affiliation{School of Physics, Institute for Research in Fundamental Sciences (IPM), Tehran 19395-5531, Iran}
\affiliation{School  of  Physics,  University  of  New  South  Wales,  Kensington,  NSW  2052,  Australia}
\affiliation{ARC Centre of Excellence in Future Low-Energy Electronics Technologies, UNSW Node, Sydney 2052, Australia}

\date{\today}

\begin{abstract}
The quasiparticle band-gap renormalization induced by the doped carriers is an important and well-known feature in two-dimensional semiconductors, including transition-metal dichalcogenides (TMDs), and it is of both theoretical and practical interest. To get a quantitative understanding of this effect, here we calculate the quasiparticle band-gap renormalization of the electron-doped monolayer MoS$_2$, a prototypical member of TMDs. The many-body electron-electron interaction induced renormalization of the self-energy is found within the random phase approximation and to account for the quasi-2D character of the Coulomb interaction in this system a Keldysh-type interaction with a nonlocal dielectric constant is used. Considering the renormalization of both the valence and the conduction bands, our calculations reveal a large and nonlinear band-gap renormalization upon adding free carriers to the conduction band. We find a 410 meV reduction of the band gap for the monolayer MoS$_2$ on SiO$_2$ substrate at the free carrier density $n=4.9\times 10^{12} \rm{cm^{-2}}$ which is in excellent agreement with available experimental results. We also discuss the role of exchange and correlation parts of the self-energy on the overall band-gap renormalization of the system. The strong dependence of the band-gap renormalization on the surrounding dielectric environment is also demonstrated in this work, and a much larger shrinkage of the band gap is predicted for the freestanding monolayer MoS$_2$.
\end{abstract}

\maketitle

\section{Introduction}
The quasiparticle band gap, defined as the energy gap between the quasiparticle conduction and valence bands, is a fundamental feature in semiconductor physics and also an important parameter in design and fabrication of electronic and optoelectronic devices such as transistors and solar cells. In this regard, the ability to tune the quasiparticle band gap via external factors can be of great importance from both theoretical and practical aspects. Among several possibilities of external manipulation in semiconductors, creating free carriers is the most common one which can be achieved by doping or generating photoexcited carriers in the system. The presence of these extra carriers not only changes the phase-space filling of the system and increases the number of interactions, but also it can affect the interaction itself through the modifications made in the screening. The overall outcome of these effects is a renormalization in the quasiparticle band gap of the system. This exchange-correlation induced renormalization of the intrinsic band gap of semiconductors is the so-called band-gap renormalization (BGR) effect.

Beyond evoking a great experimental interest, studying the band-gap renormalization and its dependence on carrier density has also been a long-standing and challenging many-body problem. The attempts to appreciate the nature of the band-gap shrinkage began in mid-1960s by studying the bulk semiconductors~\cite{inkson1976effect,abram1978heavily,mahan1980energy,berggren1981band,sernelius1986band,wagner1988band,kalt1992band} and it was continued by exploring the renormalization process in quantum wells and semiconductor heterostructures as the primary platforms of the 2D and quasi-2D electron liquid~\cite{kleinman1985band,sarma1990band,cingolani1990observation,von1992plasmaron,ryan1993band,PhysRevB.78.075211}. In these systems, the addition of free carriers and the subsequent interaction modification can renormalize the quasiparticle band gap of maximally tens of meV. A considerably larger band-gap shrinkage of about several hundred meV was also detected in one-dimensional semiconducting carbon nanotubes~\cite{spataru2010tunable}.

In the past couple of years, there has been a growing interest in transition-metal dichalcogenide (TMD) monolayers as a new realization of 2D semiconductors~\cite{mak2010atomically,splendiani2010emerging,radisavljevic2011single,mak2012control,cao2012valley,mak2013tightly}. Monolayer MoS$_2$ as a prototypical member of this vast class of materials has been the focus of huge amounts of research for many years. Beside several interesting and distinguished properties of this system, its quasiparticle and many-body features are of both fundamental and practical interest~\cite{huser2013dielectric,qiu2016screening,huser2013dielectric,PhysRevB.91.235301,thygesen2017calculating,van2019probing}. Poor screening and large Coulomb interaction, as well as the possibility of external manipulation of quantum properties, have made monolayer MoS$_2$ a potential candidate for surveying the quantum and many-body problems. Exploring the exact band gap and band-gap renormalization in this system is one of these challenging problems which has gained attention in the past few years.      

Several experimental and theoretical attempts (mostly based on density functional simulations) have been made to measure or predict the band-gap renormalization in TMDs including monolayer MoS$_2$ and to investigate the role of the doped or photoexcited carriers in this process~\cite{ugeda2014giant,chernikov2015population,qiu2019giant,liang2015carrier,qiu2016screening,gao2017renormalization,meckbach2018giant,liu2019direct}. All of these studies agree in an exceptionally large renormalization of the gap upon adding free carriers to the system. On the other hand, it is now evident that the size of the band gap in TMDs is strongly sensitive to the dielectric features of the surrounding medium~\cite{komsa2012effects,ugeda2014giant,ryou2016monolayer,raja2017coulomb,meckbach2018giant}. This is an interesting feature for practical purposes such that choosing the appropriate environment, we can have the desired band gap. Although the role of the environment on the intrinsic size of the band gap has been widely studied using optical proprieties and the binding energy of the excitons, its effect on the band-gap renormalization of the doped system has not been much explored. 

In our previous paper, we studied some many-body properties of the monolayer MoS$_2$ such as quasiparticle energy, spectral function, renormalization constant, and renormalized effective Fermi velocity, and we discussed the impact of external variables on these quasiparticle features~\cite{faridi2020many}. Pursuing the same theoretical procedure here, we focus specifically on the quasiparticle band-gap renormalization as one of the most important and practical properties of the system. Considering the important roles of the doped carriers and also the dielectric medium, we present a theoretical investigation of the band-gap renormalization in monolayer MoS$_2$ at zero temperature over a wide range of densities using the $G_0W$ approximation in the presence of various dielectric environments. Following our previous study, the dynamical screening of the carriers is considered within the many-body random phase approximation (RPA) which is a more accurate approximation in comparison with the simple plasmon-pole approximation, and furthermore, it is exact for remarkably large charge density~\cite{giuliani2005quantum}. The effect of the environmental screening is also considered assuming different substrates for the monolayer. On the other hand, to mimic a more realistic model, we have captured the effect of the out-of-plane extension of the monolayer on the band-gap renormalization using a modified Coulomb interaction of Keldysh type with a nonlocal dielectric screening~\cite{keldysh1979coulomb,cudazzo2011dielectric,torbatian2018plasmonic}. We have previously shown the crucial impact of this nonlocal screening on the quasiparticle properties of the monolayer MoS$_2$ ~\cite{faridi2020many} which also holds in the case of band-gap shrinkage. Our calculations reveal that the electron-electron interaction in the presence of free carriers decreases the quasiparticle band gap by more than 400 meV for $n=4.9\times 10^{12} \rm{cm^{-2}}$ which is in an excellent agreement with recent experimental findings~\cite{liu2019direct}. We also discuss the important effect of different substrates on band-gap renormalization in this system.              

The paper is organized as follows. In Sec. II, we present the theoretical framework including a brief introduction to the effective
Hamiltonian and the Keldysh-type electron-electron interaction. We also provide the theoretical formulation of the RPA-based self-energy and band-gap renormalization calculation. In Sec. III the numerical results of the band-gap renormalization are presented and discussed and we compare our findings with the available experimental and previous theoretical predictions. Finally, in Sec. IV we provide a brief summary.

\section{Theoretical Formulation}
In Ref.~\cite{faridi2020many} we have explained in more detail the effective low-energy Hamiltonian and the quasi-2D Coulomb interaction used to explore the quasiparticle properties of the monolayer $\rm{MoS_2}$.  For the sake of completeness, we briefly review the key points here. In order to model our electron-doped monolayer of $\rm{MoS_2}$ we use a minimal two-band Hamiltonian of the massive Dirac fermions~\cite{xiao2012coupled}. The energy dispersion of the carriers in the conduction and valence bands is then given by ${E_ k^s=s\sqrt{(\hbar v_F k)^2+\Delta^2}}$. Here $v_{\rm F}\approx5.33\times10^{5}~\rm{m/s}$ is the Fermi velocity, $2\Delta=2.7 \, \rm{eV}$ is the electronic energy gap between the valence ($s=-$) and conduction bands ($s=+$) which is the value predicted by {\it ab initio} calculations and experimental measurements~\cite{yao2017optically,gao2017renormalization}. The Fermi wave vector is also defined as $k_F=\sqrt{4\pi n/g}$ where $n$ is the carrier density and $g$ is the band degeneracy factor. For monolayer MoS$_2$ we have $g=g_s g_v=4$ where $g_s=2$ accounts for the spin degeneracy and $g_v=2$ is related to the valley degeneracy of the system.  In this work we ignore the spin splitting of the valence band which is much smaller than the electronic band gap and does not have a significant effect on the quasiparticle properties of the system.

Considering a pristine system with no electron-impurity scattering and neglecting the electron-phonon interaction, the electrons in the conduction and valence bands interact with each other through Coulomb interaction. In our calculations the intervalley interactions are ignored and to consider the important effect of the out-of-plane extension of the monolayer and the subsequent nonlocal screening of the dielectric environment, we use a Coulomb interaction of Keldysh type~\cite{keldysh1979coulomb,cudazzo2011dielectric,torbatian2018plasmonic}
\begin{equation}\label{vq}
V(q,a)=\frac{2\pi e^2}{\epsilon(q+aq^2)},
\end{equation}
where  $\epsilon=(\epsilon_1+\epsilon_2)/2$ is the average dielectric constant of the environment and $a$ is a characteristic length related to the polarizability of the 2D layer and it depends on $\epsilon$ through $a=36/\epsilon$~\AA~\cite{qiu2016screening,zhang2014absorption}. Fortunately, due to weak van der Waals interaction between monolayer MoS$_2$ and substrate, the strain induced by the lattice mismatch is not noticeable here and we can consider a system without strain~\cite{buscema2014effect,li2014photoluminescence,singh2015al2o3}. 

To calculate the band-gap renormalization (BGR), we should know how the interaction changes the conduction- and valence-band edges of the system. In an interacting system, the quasiparticle energy in band $s$ in the on-shell approximation is  given by~\cite{giuliani2005quantum}
\begin{equation}\label{E}
\mathcal{E}_Q^s(\bm k)\simeq \xi_{\bm k}^s+\rm{Re}\,{\Sigma}_s(\pmb{k},\xi_{\pmb k}^s),
\end{equation}
where ${\xi_{\bm k}^s=E_{\bm k}^s-E_{\rm F}}$ is the noninteracting energy measured from the Fermi level and ${\Sigma}_s(\pmb{k},\xi_{\pmb k}^s)$ is the self-energy of band $s$ associated with the electron-electron interaction. In this system and at zero temperature, $T=0$, the retarded self-energy of the homogeneous electron liquid of band $s$ within the $G_0W$ approximation is given by~\cite{giuliani2005quantum,mahan2013many}
\begin{equation}\label{self1}
 \begin{split}
&\Sigma_s(\bm k,\omega)=  \\
&-\sum_{s'}\int\frac{d^2\bm q}{(2\pi)^2}F^{ss'}_{\bm k,\bm k+\bm q}\int_{-\infty}^{\infty}\frac{d\Omega}{2\pi i}\frac{V_q}{\epsilon(\bm q,\Omega)}G_{0 s'}(\bm k+\bm q,\omega+\Omega),
 \end{split}
\end{equation}
where $G_{0 {s}}$ is the noninteracting Green's function of the system, $V_q$ is the short form of $V(q,a)$ given by Eq.~\eqref{vq}, ${\epsilon(\bm q,\Omega)=1-V_q\chi^0(q,\Omega)}$ is the dynamical dielectric function within the RPA, $\chi^0(\bm q,\Omega)$ is the noninteracting polarization function of the system~\cite{pyatkovskiy2008dynamical}, and $F^{ss'}_{\bm k,\bm k+\bm q}$ is the wave function overlap factor of the states $s$ and $s'$~\cite{qaiumzadeh2009effect}. 

The self-energy in Eq.~\eqref{self1} consists of a static exchange or Hartree-Fock term $\Sigma_s^{\rm{ex}}(\bm k)$ and a dynamical correlation part $\Sigma_s^{\rm cor}(\bm k,\omega)$. A formal way of decomposing the correlation self-energy is the standard line-residue decomposition which has been explained in more detail in Ref.~\cite{faridi2020many}. The total self-energy of the system is then given by
\begin{equation}
\Sigma_s(\bm k,\omega) =\Sigma_s^{\rm{ex}}(\bm k) +\Sigma_s^{\rm{line}}(\bm k,\omega) +\Sigma_s^{\rm res}(\bm k,\omega) ,
\end{equation}
where
\begin{equation}\label{ex}
\Sigma_s^{\rm{ex}}(\bm k,\omega)= -\sum_{s'}\int\frac{d^2\bm q}{(2\pi)^2}V_q F^{ss'}_{\bm{k},\bm k+\bm q} \Theta(-\xi^{s'}_{\bm k +\bm q})
\end{equation}
is the exchange part and for the correlation part we have
\begin{equation}\label{line1}
\begin{split}
\Sigma_s^{\rm line}(\bm k,\omega)=&-\negthickspace\sum_{s'}\negthickspace\int\negthickspace\frac{d^2\bm q}{(2\pi)^2}V_qF^{ss'}_{\bm k,\bm k+\bm q}\\
&\times\negthickspace\int_{-\infty}^{\infty}\negthickspace\frac{d\Omega}{2\pi}\Bigl[\frac{1}{\epsilon(\bm q,i\Omega)}\negthickspace-1\negmedspace\Bigr]\frac{1}{\omega\negmedspace+\negmedspace i\Omega\negmedspace-\negmedspace\xi^{s'}_{\bm k+\bm q}},
\end{split}
\end{equation}
and 
\begin{equation}\label{res}
\begin{split}
\Sigma_s^{\rm res}(\bm k,\omega)=&\sum_{s'}\int\frac{d^2\bm q}{(2\pi)^2}V_q\Bigl[\frac{1}{\epsilon(\bm q,\omega-\xi_{s'}(\bm k+\bm q))}-1\Bigr]\\
&\times F^{ss'}_{\bm k,\bm k+\bm q}\bigr[\Theta(\omega-\xi^{s'}_{\bm k+\bm q})-\Theta(-\xi^{s'}_{\bm k+\bm q})\bigl].
\end{split}
\end{equation}
We can see that each of these terms also contains the interband and intraband self-energies which result from considering both interband and intraband interactions between electrons. Considering ${k}=0$ in Eq.~\eqref{E} for $s=+$ and $s=-$, the interaction-induced shift of the conduction-band minimum and the valence-band maximum can be found.
It should be noted that before populating the conduction band, the valence band has already been full and since we are always interested in the changes of the self-energy, the self-energy of the undoped system should be subtracted from its doped counterpart~\cite{inkson1976effect,berggren1981band}. Therefore, the quasiparticle dispersion of the conduction-band minimum is given by
\begin{equation}\label{cb}
\mathcal{E}_Q^c(\bm k=0)\simeq \xi_{0}^++\rm{Re}\,{\widetilde{\Sigma}}_+(0,\xi_{0}^+),
\end{equation}
with 
\begin{equation}\label{cb}
\widetilde{\Sigma}_+(0,\xi_{0}^+)=\Sigma_+(0,\xi_{0}^+)-\Sigma_+^{\rm{ex,inter}}(0),
\end{equation}
where the second term on the right-hand side is the self-energy of the conduction band for the undoped system which is an interband exchange term~\cite{hwang2007density} and is found from  Eq.~\eqref{ex} putting $s=+$ and $s'=-$.

In the same way, for the valence-band maximum we have
\begin{equation}\label{cb}
\mathcal{E}_Q^v(\bm k=0)\simeq \xi_{0}^-+\rm{Re}\,{\widetilde{\Sigma}}_-(0,\xi_{0}^-),
\end{equation}
with 
\begin{equation}\label{cb}
\widetilde{\Sigma}_-(0,\xi_{0}^-)=\Sigma_-(0,\xi_{0}^-)-\Sigma_{-}^{\rm{ex,intra}}(0),
\end{equation}
where the self-energy of the valence-band edge for the undoped system is given by an intraband Hartree-Fock term considering the $s=-$ and $s'=-$ case in Eq.~\eqref{ex}.

Finally the electron-electron induced BGR is given by
\begin{equation}\label{cb}
\rm{BGR}=\Delta E_g=\rm{Re}\,{\widetilde{\Sigma}}_+(0,\xi_{0}^+)-\rm{Re}\,{\widetilde{\Sigma}}_-(0,\xi_{0}^-),
\end{equation}
where the first term on the right-hand side shows the conduction-band minimum renormalization (CBR) while the second term gives the renormalization of the valence-band maximum (VBR). It is important to note that to avoid the wrong and inaccurate numerical results for the BGR, the integrals in Eqs.~\eqref{ex}-\eqref{res} should be rewritten for the special case $\bm k=0$.

\section{band-gap renormalization results}

\begin{figure}[h]
\centering
  \includegraphics[width=1.\linewidth]{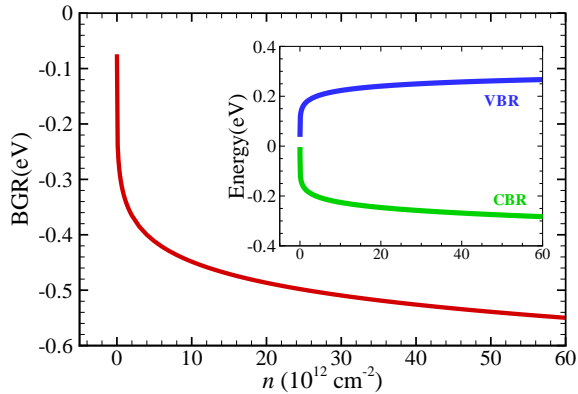}
\caption{\label{fig1} (Color online) The band-gap renormalization (in units of eV) of doped monolayer $\rm{MoS_2}$ on $\rm{SiO_2}$ substrate with effective dielectric constant $\epsilon=2.5$ as a function of the electron doping density (in units of 10$^{12}$  cm$^{-2}$). Inset: The conduction-band minimum and valence-band maximum renormalization (CBR and VBR) as functions of the doping density. Notice that the intraband interactions in the conduction band and the interband interactions in the valence band grow upon adding free carriers; however, the former is greater than the latter owing to the large gap between the bands.
    }
\end{figure}

In Fig. \ref{fig1} we show our results for the total BGR of  an electron-doped monolayer $\rm{MoS_2}$ on $\rm{SiO_2}$ substrate with effective dielectric constant $\epsilon=2.5$ as a function of doping density. 
A very large band-gap renormalization of about $0.55~ \rm{eV}$ is achieved when the density is increased up to $n=6\times 10^{13} \rm{cm^{-2}}$. The most significant part of this nonlinear and large renormalization belongs to the low-density region such that about $70\%$ of the band-gap shrinkage occurs with $n=3\times 10^{12} \rm{cm^{-2}}$ showing that even a light doping can strongly decrease the band gap of a monolayer $\rm{MoS_2}$. In the inset, we show separately the conduction-band (CBR) and the valence-band (VBR) contributions in the total band-gap renormalization. We can see that the interaction leads to a downward shift of the conduction-band minimum and at the same time it causes an approximately identical upward shift in the valence-band maximum. It turns out that adding free carriers to the conduction band leads to a reduction of the interaction in the valence band and finally results in an upward shift in this band. As the doping density is increased, the conduction-band renormalization starts to exceed the renormalization in the valence band. This is because upon adding free carriers in the conduction band, the intraband interactions in this band and the interband interactions in the valence band grow, but due to the large gap between the bands, the former is greater than the latter. Our results of the band-gap renormalization calculation are in good agreement with the experimental findings of Ref.~\cite{liu2019direct}. In that paper, the authors report a band-gap reduction of about $0.4~\rm{eV}$ for an $n$-doped monolayer $\rm{MoS_2}$ on $\rm{SiO_2}$ substrate at $n=4.9\times 10^{12} \rm{cm^{-2}}$ (before optical excitation). At the same doping, our calculations show a $0.41~\rm{eV}$ band-gap renormalization which obviously indicates a very good quantitative agreement with the measured reduction of the band gap. 
\begin{figure}[h]
\centering
  \includegraphics[width=1.\linewidth]{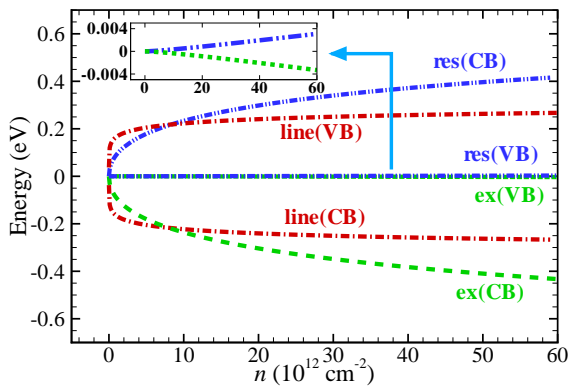}
\caption{\label{fig2} (Color online) Exchange (ex), residue (res), and line contributions of the total self-energy in conduction- and valence-band renormalization as functions of the doping density. Inset: The enlarged plot of the residue and exchange terms in the valence band.
    }
\end{figure}

\begin{table*}[htb]
\centering
\renewcommand\arraystretch{1.3}
\caption{Numerical values of the conduction- and valence-band renormalization (CBR and VBR) and the exchange (ex), residue (res), and line contributions of the total self-energy in monolayer MoS$_2$.}\label{tab1}
\begin{tabular}{@{\extracolsep{9pt}}ccccccccc}
\\[-1.8ex]\hline\hline\\[-1.8ex]
{} & {} & Conduction band & {}&{} & {} & Valence band &{}& {} \\
\cline{2-5} \cline{6-9} \\[-1.2ex] 
$n\,(10^{12}~\rm{cm^{-2}})$ & ex(eV)  & res(eV)  & line(eV)  & CBR(eV)& ex($10^{-3}$eV)  & res($10^{-3}$eV)  &line(eV)& VBR(eV) \\
\hline\\[-1.8ex]
$ 1 $ &-0.091 & 0.091 &-0.166 &-0.166& -0.01& 0.01 & 0.166&0.166\\
$5$ & -0.180 & 0.179 &-0.205&-0.207& -0.14& 0.14 & 0.205&0.205\\
\\[-1.8ex]
$ 10 $ &-0.236 &0.233 &-0.223 &-0.226& -0.36& 0.35 & 0.223&0.223\\
$20$ &-0.304& 0.298 & -0.241 &-0.246&-0.87& 0.84 & 0.241&0.240\\
\\[-1.8ex]
$ 40 $ &-0.382& 0.371 & -0.258 &0.269& -2.05& 1.94 & 0.258&0.258\\
$60$ &-0.433& 0.417 &-0.267&-0.283&-3.31& 3.09 & 0.267&0.267\\
\hline

\end{tabular}
\end{table*}

To perceive the impact of many-body electron-electron interaction, the contributions of different terms of the total self-energy are illustrated separately in Fig.~\ref{fig2} for both the conduction-band minimum and the valence-band maximum. Note that here the exchange self-energy of the conduction band contains only the intraband term while that of the valence band shows the interband term. We can see that in the conduction band the negative exchange self-energy tends to reduce the band-edge energy through altering the electron occupation. On the other hand, the overall correlation self-energy [$\Sigma_+^{\rm{line}}+\Sigma_+^{\rm res}$] being negative in a low-density regime helps in sharper reduction of the conduction-band minimum. As the density is increased this term changes sign and grows gradually with the density and decelerates the reduction of the band-edge energy in the high-density regime. 

Looking at the correlation parts separately, it turns out that the absolute value of $\Sigma_+^{\rm{line}}$ which expresses the effect of carrier screening on the self-energy of the system has an abrupt decrease at low densities (reducing the electron-electron interaction) and a saturating behavior at high densities. The $\Sigma_+^{\rm{res}}$ term is also a density-dependent term and at the same time the changes in phase-space filling affect this term. At the conduction-band edge, the interband interactions have no contribution in the residue term, and we can see that this term is very similar to the exchange self-energy but with the opposite sign especially at low densities. In fact, the combination of the residue and the exchange self-energies at the band edge is approximately equal to the so-called screened exchange self-energy which grows linearly with the density and is much weaker than the unscreened exchange self-energy. In the case of the valence band, the line term is the same as that of the conduction band but with the opposite sign which shows that the screening of the free carriers weakens the interaction of the electrons in the valence band and causes an upward shift in the energy of the band-edge quasiparticles. Contrary to the conduction band, the residue term here only contains an interband term and that is why both the residue and the exchange self-energies are very small. In fact the poor overlap between the wave functions of the valence and the conduction band reduces the value of these terms. 

The overall behavior of the exchange and correlation parts of the self-energy indicates that at low density, the residue and exchange terms are approximately equal but with opposite signs and therefore the line part of the correlation self-energy of the two bands determines the sharp shrinkage of the band-gap renormalization in this regime. However, upon increasing the free carrier density, both the saturating character of the line term and the increasing value of $\Sigma^{\rm{ex}}+\Sigma^{\rm{res}}$ play a role in the ultimate evolution of the band-gap renormalization in this regime. 

As the behavior of the BGR in low densities is determined by the line part of the correlation self-energy, we can conclude that the carrier screening is the dominant mechanism in this limit. It has long been known that the BGR in 2D and quasi-2D semiconductors is larger than their three-dimensional counterparts~\cite{sarma1990band,cingolani1990observation}. This is owing to the poor intrinsic screening and consequently larger Coulomb interaction in these systems. In this case, the system is very sensitive to additional screening mechanisms such as free-carrier induced screening or environmental screening. Therefore, the screening induced by even a low density of doped carriers in two-dimensional systems can strongly affect the interaction and reduce the band gap of the system. This is also the case when a moderate dielectric material is present in the vicinity of the monolayer $\rm{MoS_2}$ where environmental screening plays a role. In general, adding free carriers to the conduction band introduces new poles to the screened interaction (or zeros to the dielectric function), the so called plasmon excitations, which lead to additional electronic screening effect due to free carriers. This effect lies at the heart of the $\Sigma^{\rm{line}}$ and is crucial in determining its behavior. Unfortunately we can not find an analytical expression for the asymptotic behavior of the BGR for the low-doping regime. This is due to the complications associated with momentum and frequency integration of  $\Sigma^{\rm{line}}$ and also the complexity of the dielectric function of the system calculated in the RPA. Meanwhile our numerical results show an $n^{1/7}$ behavior for the band-gap renormalization at the low-doping regime with carrier density in the range $10^{10}-10^{12}$ cm$^{-2}$. As the density is increased both the saturating character of the line part and the positive increase of $\Sigma^{\rm{ex}}+\Sigma^{\rm{res}}$ lead to a smoother behavior of the BGR (approximately $n^{1/9}$ for $2\times10^{13}$ $<n<6\times10^{13}$ cm$^{-2}$). 
We have summarized some of our results for the conduction- and valence-band renormalization (CBR and VBR) and the discussed contributions of the total self-energy in Table \ref{tab1}.

\begin{figure}[h]
\centering
  \includegraphics[width=1.\linewidth]{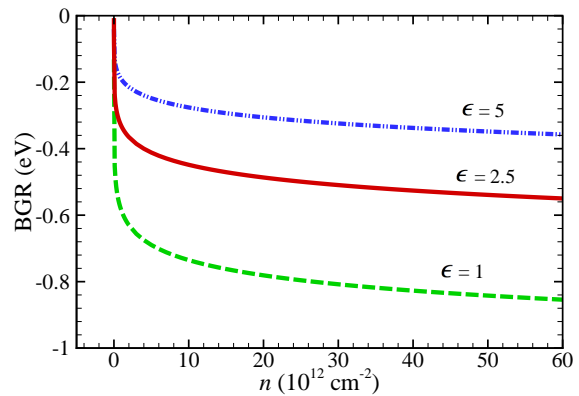}
\caption{\label{fig3} (Color online) Band-gap renormalization of doped monolayer $\rm{MoS_2}$ on $\rm{Al_2O_3}$ with $\epsilon=5$, $\rm{SiO_2}$ with $\epsilon=2.5$, and suspended monolayer $\rm{MoS_2}$ with $\epsilon=1$ as functions of doping density. Notice that a small dielectric constant can lead to a large band-gap renormalization such that for a freestanding monolayer $\rm{MoS_2}$, significant band-gap shrinkage (about  0.6 eV) occurs with a density as low as $n=1.2\times 10^{12} \rm{cm^{-2}}$.  }
\end{figure}
 
The effect of the substrate is also of great importance in the results of the band-gap renormalization due to the vast impact of the environmental screening on the strength of the Coulomb interaction. In Fig. \ref{fig3} we have compared the BGR results of monolayer $\rm{MoS_2}$ on two different substrates: $\rm{SiO_2}$ with effective dielectric constant $\epsilon=2.5$ and $\rm{Al_2O_3}$ with $\epsilon=5$ and also the suspended monolayer $\rm{MoS_2}$ with $\epsilon=1$. We can see that in comparison with the $\epsilon=2.5$ case with $\rm{BGR}\approx 0.55~\rm{eV}$, the band-gap renormalization reduces to 0.36 eV for $\epsilon=5$ while for a monolayer $\rm{MoS_2}$ with no substrate the band-gap shrinkage is as large as 0.85 eV (for $n=6\times 10^{13} \rm{cm^{-2}}$). This effect has also been discussed in Ref.~\cite{meckbach2018giant} for a photoexcited monolayer $\rm{MoS_2}$ and although the treatment in doped and photoexcited systems is not exactly the same, our findings are remarkably similar for the suspended monolayer $\rm{MoS_2}$ and also $\rm{MoS_2}$ on $\rm{SiO_2}$. We can also realize from this figure that even a light doping in a system surrounded by a medium with a small dielectric constant can lead to a large band-gap renormalization such that for a freestanding monolayer $\rm{MoS_2}$, $70\%$ of the band-gap shrinkage ($\approx$ 0.6 eV) occurs with a density as low as $n=1.2\times 10^{12} \rm{cm^{-2}}$. Before concluding this part, we want to mention that even in the case of undoped $\rm{MoS_2}$, the dielectric constant of the surrounding medium plays a crucial role in the intrinsic band gap of the system such that the band gap of the undoped monolayer $\rm{MoS_2}$ varies from 1.8 eV to 2.8 eV depending on the environment~\cite{ryou2016monolayer}. Since the intrinsic band gap is an input parameter in our calculations, we investigate its impact on the BGR and find that changing the intrinsic band gap from 1.8 eV to 2.8 eV does not cause a noticeable change in the BGR. Meanwhile the system surrounded by a smaller dielectric environment and higher densities is comparably more affected such that for a freestanding monolayer $\rm{MoS_2}$, the change in the calculated BGR is about 18 meV in the low-density regime, $n=10^{12} \rm{cm^{-2}}$ and 34 meV for $10^{13} \rm{cm^{-2}}$. In the case of monolayer $\rm{MoS_2}$ on a substrate with $\epsilon=5$ and for the same densities, the change in the BGR is not more than 9 and 21 meV, respectively.

\section{Summary}
To summarize, we have obtained the quasiparticle band-gap renormalization of the electron-doped monolayer MoS$_2$ within G$_0$W and the RPA. A large and nonlinear renormalization of the band gap is found considering the contributions of both conduction and valence bands. We have shown that upon adding free carriers to the conduction band, an upward shift in the valence-band edge together with an inverse downward shift in the conduction-band minimum result in the overall renormalization of the band gap.
We have also discussed the contributions from the exchange and correlation parts of the self-energy on the valence- and conduction-band renormalization in low- and high-density regimes. We should emphasize that considering the nonlocal dielectric screening in this system through a modified Coulomb interaction is absolutely crucial in obtaining the final results, such that ignoring this point we would find extremely large and nonphysical values for the BGR due to exceptionally strong interactions.

Finally, the important and less-studied effect of the environmental dielectric medium has been considered in this work, and we have found that the substrate-induced screening has a major effect on the quasiparticle band-gap renormalization in monolayer MoS$_2$ such that a medium with smaller effective dielectric constant gives rise to a much larger band-gap renormalization in this system.
Our results agree well with recent experimental measurements and previous theoretical findings and can pave the way for understanding the combined effect of doping and dielectric medium on the band-gap renormalization of the monolayer MoS$_2$.

This approach can also be generalized to calculate the renormalized spin-orbit coupling in transition-metal dichalcogenide monolayers for a hole-doped case. An improvement to this study is considering the finite-temperature impact on BGR calculation. Qualitatively, we expect a decrease in BGR as the temperature is increased. This is owing to the fact that as we increase the temperature, the decaying channel of quasiparticles into plasmons starts to wipe out and consequently the contribution of plasmons in the dielectric function of quasiparticles is washed out. Since this contribution plays a major role especially in the line part of the self-energy, a smaller renormalization in the band gap is expected at finite temperature. But the quantitative estimation of the BGR and its behavior at different doping regimes needs an exact finite-temperature self-energy calculation.    

\section{Acknowledgement}
D. C. and R. A. were supported by the Australian Research Council Centre of Excellence in Future Low-Energy Electronics Technologies (Project No. CE170100039).

\end{document}